# The (impossible?) formation of acetaldehyde on the grain surfaces: insights from quantum chemical calculations

**J. Enrique-Romero,[1] A. Rimola,[1]\* C. Ceccarelli[2,3] and N. Balucani[2,3,4]**

[1]*Departament de Quimica, Universitat Autonoma de Barcelona, 08193, Bellaterra, \*E-mail: albert.rimola@uab.cat*
[2]*IPAG, Université Grenoble Alpes, F-38000 Grenoble, France*
[3]*CNRS, IPAG, F-38000 Grenoble, France*
[4]*Dipartimento di Chimica, Biologia e Biotecnologie, Università di Perugia, Via Elce di Sotto 8, I-06123 Perugia, Italy*

**ABSTRACT**

Complex Organic Molecules (COMs) have been detected in the interstellar medium (ISM). However, it is not clear whether their synthesis occurs on the icy surfaces of interstellar grains or via a series of gas-phase reactions. As a test case of the COMs synthesis in the ISM, we present new quantum chemical calculations on the formation of acetaldehyde ($CH_3CHO$) from the coupling of the HCO and $CH_3$ radicals, both in gas phase and on water ice surfaces. The binding energies of HCO and $CH_3$ on the amorphous water ice were also computed (2333 and 734 K, respectively). Results indicate that, in gas phase, the products could be either $CH_3CHO$, $CH_4 + CO$, or $CH_3OCH$, depending on the relative orientation of the two radicals. However, on the amorphous water ice, only the $CH_4 + CO$ product is possible due to the geometrical constraints imposed by the water ice surface. Therefore, acetaldehyde cannot be synthesized by the $CH_3$ + HCO coupling on the icy grains. We discuss the implications of these results and other cases, such as ethylene glycol and dimethyl ether, in which similar situations can occur, suggesting that formation of these molecules on the grain surfaces might be unlikely.

## 1 INTRODUCTION

Of the almost 200 molecular species detected in the interstellar medium (ISM), all those containing more than six atoms bear at least one carbon atom. Thus, understanding organic chemistry in the ISM is mandatory if we want to understand interstellar chemistry at all. Moreover, the organic chemistry occurring in the ISM can reach a certain degree of complexity, up to organic compounds of prebiotic interest. Indeed, knowing whether the organic molecules synthesized in space contributed or not to the chemical evolution needed for the emergence of life on Earth is one of the open questions in the origin of life studies.

The presence of complex, from an astronomical point of view, organic molecules (hereinafter we will call them COMs, Complex Organic Molecules) in the ISM has been known for decades, specifically in hot cores and corinos (e.g. Blake et al. 1986; Cazaux et al. 2003). Lately, COMs have also been detected in cold gas, about 10 K (Bacmann et al. 2012; Cernicharo et al. 2012; Vastel et al. 2014). Two classes of models are invoked to explain the synthesis of COMs. The first one, and nowadays most popular, postulates that COMs are synthesized on the grain surfaces in a three-step process (Garrod & Herbst 2006; Garrod, Weaver & Herbst 2008; Öberg et al. 2009; Kalvans 2015): (1) formation of hydrogenated species, like $CH_3OH$ or $H_2CO$, by H-addition(s) to frozen atoms/molecules, like O or CO, during the cold pre-collapse phase; (2) formation of radicals, like $CH_3O$ or HCO, from the dissociation of the frozen hydrogenated species by the UV photons caused by the interaction of the cosmic rays with the molecular hydrogen of the pre-collapse core; (3) coupling of radicals on to COMs on the surfaces when the grain temperature reaches ∼30 K and the radicals, therefore, acquire mobility and diffuse on the surface, during the collapse phase. Recently, Ruaud et al. (2015) proposed a set of reactions on the grain surfaces induced by carbon atoms, which are particularly abundant at early times in the molecular cloud evolution, always following the idea that eventually radicals combine together to form some COMs in cold environments. The second class of models, which was in vogue before 2006 (e.g. Charnley, Tielens & Millar 1992; Caselli, Hasegawa & Herbst 1993) and has been revived in 2015 (e.g. Balucani, Ceccarelli & Taquet 2015; Barone et al. 2015; Taquet, Wirström S & Charnley 2016), postulates that COMs are formed via gas-phase reactions by a two-step process: (1) same as that in the grain-surface models; (2) the hydrogenated species of the pre-collapse phase are released into the gas phase, by either thermal evaporation when the grain temperature exceeds ∼100 K (hot cores/corinos) or by photo-desorption from interstellar UV photons, and react with other gaseous species to form COMs through a series of gas-phase processes. In practice, in both models, the precursor molecules are hydrogenated molecules that are synthesized on the grain surfaces by addition of Hatoms, such as the case of methanol (Tielens & Hagen 1982;

Watanabe & Kouchi 2002; Taquet, Ceccarelli & Kahane 2012; Rimola et al. 2014). The two classes of models, however, differ in what happens to those hydrogenated species; that is, how they are 'cooked' to synthesize COMs.

The problem in understanding which of the two paradigms, grain-surface versus gas-phase COMs synthesis, is correct and in which conditions each paradigm is applicable mostly comes from the dramatic lack of experimental and/or theoretical data on the reactions in gas phase and on grain surfaces. In this Letter, we focus on the acetaldehyde ($CH_3CHO$) synthesis, which can be considered a test case for the more general problem of COMs formation in the ISM. This molecule has been detected in several environments, from cold (10 K) objects (Öberg et al. 2010; Bacmann et al. 2012; Jaber et al. 2014; Vastel et al. 2014) to hot cores/corinos (e.g. Blake et al. 1986; Cazaux et al. 2003) and molecular shocks (Codella et al. 2015). The current grain-surface models postulate that acetaldehyde forms by the coupling of $CH_3$ with HCO, previously formed by the photodissociation of methanol and formaldehyde, respectively (Garrod et al. 2008; Öberg et al. 2009), or by successive H-atom additions on CCO (Ruaud et al. 2015). Here, we focus on the former channel; i.e. $CH_3 + HCO \rightarrow CH_3CHO$, which is assumed to proceed without any energy barrier in the above mentioned models. We present new theoretical computations of this reaction both in gas phase and on an amorphous water ice surface modelled by a molecular cluster, in which the binding energies of $CH_3$ and HCO on the water ice model are also provided.

## 2 METHODS AND RESULTS

### 2.1 Methods

Molecular calculations were performed using the GAUSSIAN09 program package (Frisch et al. 2013). The structure of the molecular stationary points was fully optimized using the global hybrid M06-2X functional (Zhao & Truhlar, 2008) combined with the a posteriori Grimme's D3 correction (Grimme et al. 2010) to take dispersion interactions into account. For the gas-phase reactions, the standard 6-311++G(2df,2pd) basis set was used, whereas on the water ice model the standard 6-311++G(d,p) basis set was employed. All structures were characterized by the analytical calculation of the harmonic frequencies as minima (reactants, intermediates, and products) and saddle points (transition states). Thermochemical corrections to the potential energy values to obtain zero-point energy corrected values were carried out using the standard rigid rotor/harmonic oscillator formulas (McQuarrie, 1986).

In this work, calculations based on a periodic approach were also performed using the periodic ab initio CRYSTAL14 code (Dovesi et al. 2014). For these calculations, we used the B3LYP-D2* density functional method, which combines the B3LYP method (Lee, Yang & Parr 1988; Becke 1993) with a modification of the D2 term proposed by Grimme (2006) to account for dispersion, providing accurate cohesive (Civalleri et al. 2008) and adsorption energy values (Rimola, Civalleri & Ugliengo 2010) within a periodic treatment. The adopted Gaussian functions consisted of an all electron 6-311G(d,p) standard basis set. All calculations involving one radical species were run as open-shell systems based on the unrestricted formalism. For those involving two radical species, the starting guess were open shell broken symmetry, but all the cases collapsed to a closed shell system. Following the International System Units, all energy units are given in kJ mol$^{-1}$, whose conversion factor to K is 1 kJ mol$^{-1}$ = 120.274 K.

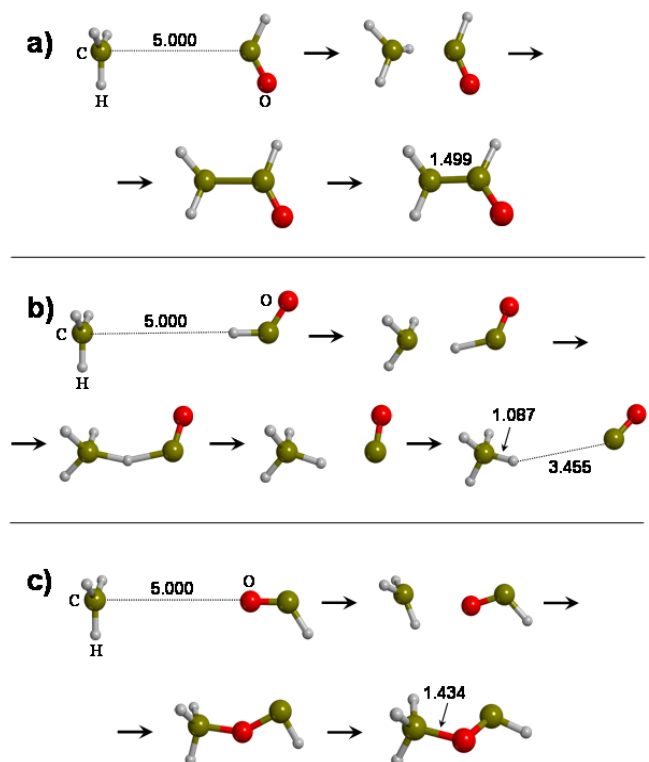

**Figure 1.** Sequence of the geometry optimization at M062X-D3/6-311++G(2df,2pd) theory level of reaction between HCO and $CH_3$ adopting the following initial guess structures: (a) the C atoms are directly pointing one to each other with an initial distance of 5 Å, (b) the H atom of HCO is pointing towards the C atom of $CH_3$ with an initial distance of 5 Å, and (c) the O atom of HCO is pointing towards the C atom of $CH_3$ with an initial distance of 5 Å. The resulting products, formed spontaneously upon geometry optimization, are $CH_3CHO$, $CH_4$ + CO, and $CH_3OCH$, respectively, whose optimized geometries are shown in the last image of the corresponding sequences. Bond distances are in Å.

### 2.2 Results

The reaction between the HCO and $CH_3$ radicals in gas-phase conditions leads to the formation of different products, depending on the relative orientation of the reacting radicals in the initial guess structures (i.e. before geometry optimization). When the C atoms of HCO and $CH_3$ are directly pointing one towards each other, the geometry optimization yields a spontaneous C–C coupling, leading to the formation of acetaldehyde (see sequence of Fig. 1a). When the H atom of HCO is pointing towards the C atom of CH$_3$, a spontaneous H transfer is given upon geometry optimization, thus forming CH$_4$ and CO as the final products (see Fig. 1b). Finally, when the O atom of HCO points towards the C atom of CH$_3$ the system collapses to CH$_3$OCH (see Fig. 1c).

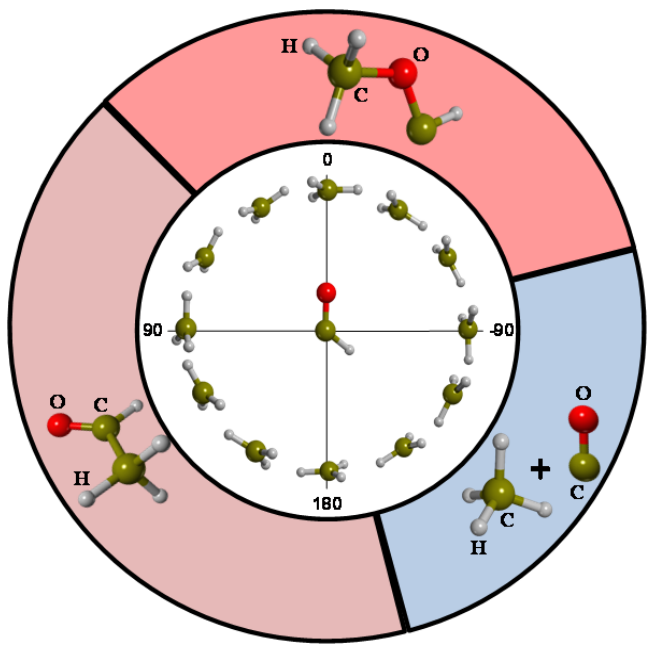

**Figure 2.** Summary of the products spontaneously formed upon geometry optimization by reaction of HCO and CH$_3$, in which the reactants exhibit different H$_3$C-C-OH angles in the initial guess structures.

**Table 1.** Reaction potential energies ($\Delta E$) and with the zero-point energy corrections ($\Delta U_0$) for the formation of CH$_3$CHO, CH$_4$ + CO, and CH$_3$OCH in gas phase, calculated at M062X-D3/6-311++G(2df,2pd) theory level. Values in units of kJ mol$^{-1}$.

| Product | $\Delta E$ | $\Delta U_0$ |
|---|---|---|
| CH3CHO | -383.3 | -349.6 |
| CH4 + CO | -388.3 | -367.4 |
| CH3OCH | -104.9 | -71.8 |

Table 1 reports the calculated reaction energies (based on pure potential energies $\Delta E$, and including zero-point energy corrections $\Delta U_0$) of these processes, considering the CH$_3$ + HCO asymptote the zero-energy reference. According to these values, formation of CH$_4$ + CO is the most exoergic process ($\Delta U_0$ = −367.4 kJ mol$^{-1}$), but it is followed (a difference of 22 kJ mol$^{-1}$ or 2650 K) by formation of CH$_3$CHO, with $\Delta U_0$ = −349.6 kJ mol$^{-1}$. Formation of CH$_3$OCH is, finally, the least favourable process ($\Delta U_0$ = −71.8 kJ mol$^{-1}$).

In order to have a deeper insight into the influence of the relative orientations of the CH$_3$ and HCO reactants, we have performed a set of geometry optimizations in which the initial guess structures exhibit different H$_3$C-C-OH angles, spanning the 0–360 deg range with variations of 30 deg. Fig. 2 summarizes the obtained results. For angles between 60 and 180 deg formation of acetaldehyde dominates; between −90 and −150 deg the reaction is governed by formation of CH$_4$ + CO; whereas between angles of 30 and −60 deg the direct product is CH$_3$OCH. All these results indicate, in summary, that the immediate product formed by the reaction of CH$_3$ with HCO strongly depends on how the radicals are oriented. Interestingly, in gas phase, the two radicals can move freely and van der Wals forces are expected to orient the two radicals as they get close. According to combustion related studies, the H-transfer channel leading to CH$_4$ + CO is largely favoured (Tsang & Hampson 1986). Moreover, although the formation of these three products is possible and the reactive channels are competitive, it is worth to highlight that the CH$_3$CHO and CH$_3$OCH products can redissociate back to reactants if they are not stabilized through three-body reactions.

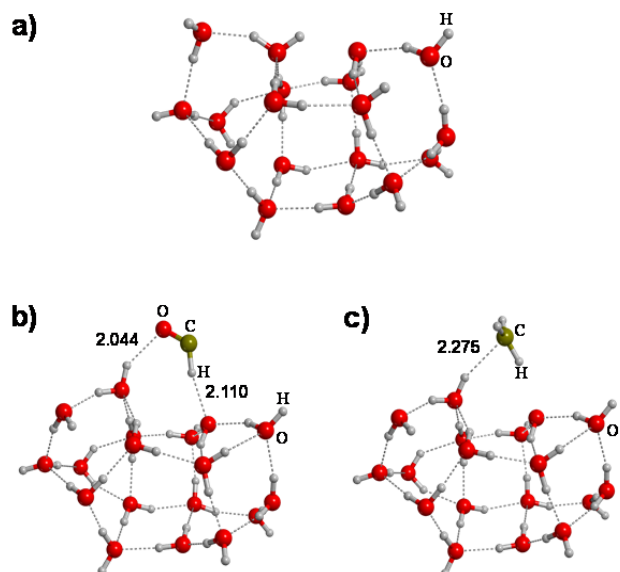

**Figure 3.** M062X-D3/6-311++G(d,p)-optimized structures for (a) the wàter ice cluster model consisting of 18 water molecules, (b) the water ice cluster interacting with HCO, and (c) the water ice cluster interacting with CH3. Bond distances are in Å.

We have also studied the very same reaction on a 18-H$_2$O clúster model representing the surface of an amorphous interstellar wàter ice (see Fig. 3a). This cluster model has also been used by some of us to simulate the successive hydrogenation of CO to form CH$_3$OH, providing very similar structural and energetic results to those using larger water ice clusters (Rimola et al. 2014). For the reaction we are concerned with, the initial guess structure is constrained by the intermolecular interactions established between the water ice cluster and the CH$_3$ and HCO reactants. The interaction of HCO alone and CH$_3$ alone with the water ice cluster has been examined by optimizing the geometry of these adducts, whose results are shown in Fig. 3(b) and Fig. 3(c), respectively. For both systems, HCO and CH$_3$ are engaged by hydrogen bond (H-bond) interactions with the water ice surface; that is, HCO through H-bonds of moderate strengths involving the H and O ends (acting as H-bond donor and acceptor, respectively), and CH$_3$ through a weak H-bond in which the C atom acts as an H-bond acceptor. As expected, the interaction of HCO is stronger than that of CH$_3$, because of the larger polarity of the former species. The calculated interaction (binding) energies including zero-point energy corrections are −19.4 and −6.1 kJ mol$^{-1}$ (2333 and 734 K), for HCO and CH$_3$, respectively. It is worth mentioning that dispersion interactions are also contributing to these values as they are accounted for in the methods used.

In the light of these results, a suitable and logical initial guess structure to carry out the coupling between HCO and CH$_3$ is that shown in the first image of Fig. 4(a), in which HCO and CH$_3$ are H-bonded with the water ice cluster. This initial structure evolves upon geometry optimization according to the sequence shown in Fig. 4(a). The H atom of HCO is spontaneously transferred to a nearby interacting water molecule of the ice, thus triggering an H-atom transfer mechanism between different water molecules and finishing with a final H transfer to CH$_3$. Therefore, the final structure collapsed on to the formation of CH$_4$ and CO. In this case, the interaction between HCO and CH$_3$ with the water ice cluster imposes an initial geometry actually prone to driving the formation of CH$_4$ and CO, hampering the formation of CH$_3$CHO through coupling of respective C atoms. In order to assess if these results are biased by the limited size of the water cluster model, we calculated the same process on a larger water cluster consisting of 32 water molecules, which was used by some of us in previous works (e.g. Duvernay et al. 2014; Rimola et al. 2014). Results indicate that the final product is also CH$_4$ + CO due to a spontaneous H-transfer process (see sequence shown in Fig. 4b).

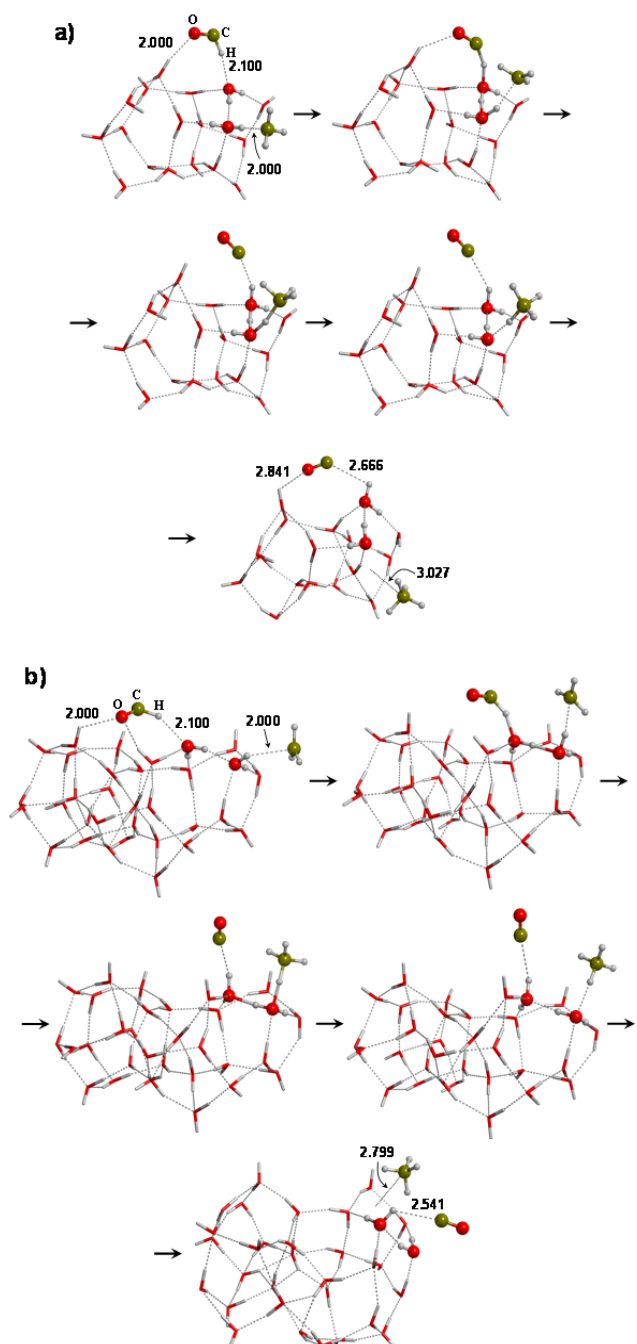

**Figure 4.** Sequence of the geometry optimization at M062X-D3/6-311++G(d,p) theory level for the HCO and CH$_3$ species on the water ice cluster models consisting of 18 (a) and 32 (b) water molecules. The initial distances are highlighted in the first image of the sequences. In both cases the systems collapse on to formation of CH$_4$ and CO. The optimized geometries are shown in the last image of the sequences. Bond distances are in Å.

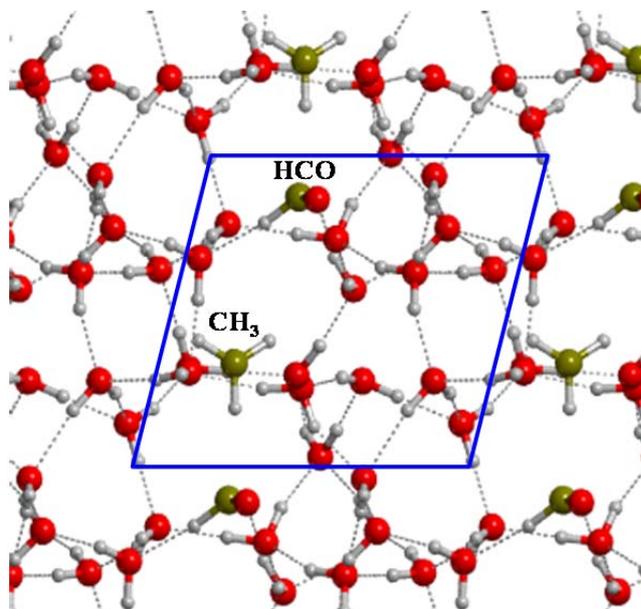

**Figure 5.** B3LYP-D2*/6-311G(d,p)-optimized structure for CH$_3$ and HCO fully embedded in the bulk of an amorphous water ice simulated by a periòdic calculation. The unit cell of the system is highlighted in blue.

It is reasonable to consider that at the ice temperature where HCO and CH3 acquire some mobility, they could also reorient in a favourable way to form CH$_3$CHO. Nevertheless, according to our results, the H-transfer mechanism takes place in a barrierless fashion, and, thus, it is spontaneous once HCO and CH$_3$ are simultaneously adsorbed on two contiguous sites of the water ice. Therefore, the formation of CH$_4$ and CO will be always the dominant process on water ices.

Finally, it is also of interest to check if the same spontaneous process can take place in the case that the radicals are inside the water ice rather than adsorbed on the surface. This is a plausible scenario since UV photons and cosmic rays can also have an impact on molecules in the sub-surface and generate radicals in the bulk ice. To have a deeper insight into this aspect, we have carried out a periodic calculation of CH$_3$ and HCO fully immersed in an amorphous water ice. Our results indicate that the system is stable; i.e. no spontaneous reactions take place (see Fig. 5). This is because the radicals are stabilized by the interactions with the surrounding water molecules. This is particularly true for the HCO radical, which is engaged in different H-bonds. Therefore, a reactive interaction of CH$_3$ and HCO leading to the spontaneous formation of CH$_4$ + CO is only possible when the radicals are adsorbed on the ice surfaces and not inside the water bulk.

## 3 DISCUSSION AND CONCLUSIONS

The aim of this work is to shed some light into the more general problem of the COMs synthesis in the ISM: where do they occur? on the surfaces of interstellar grains or in gas phase? To this end, we have carried out accurate quantum chemical calculations focused on the formation of $CH_3CHO$ from the coupling of the HCO and $CH_3$ radicals, in gas phase and on the surface of an amorphous water ice.

We have found that gaseous $CH_3$ and HCO, which are free to reorient, can undergo an H-transfer reaction (leading to $CH_4$ + CO) or couple to form $CH_3CHO$ or $CH_3OCH$. The three processes are barrierless and can, therefore, compete, although in the absence of three-body reactions the coupling products $CH_3CHO/CH_3OCH$ will redissociate back to reactants. In contrast, on a water ice surface (in this work modelled by a molecular cluster consisting of 18 water molecules), only the $CH_4$ + CO products can be formed. This exclusive formation is due to the geometrical constraints imposed by the interactions of the HCO and $CH_3$ reactants with the water ice surface (mainly H-bond and dispersion), which favour a spontaneous H-transfer from HCO to $CH_3$ through a H-transfer mechanism assisted by several water molecules of the ice. Also when enough energy is provided to the system and some mobility of the reactants on the icy surface is present, the $CH_4$ + CO channel remains the favoured one. It is worth mentioning that the experimental works where formation of $CH_3CHO$ was observed by processing icy grains (Bennett et al. 2005; Öberg et al. 2009), the initial ice mixtures did not contain water, while, according to our results, it is its presence that hampers the formation of $CH_3CHO$.

As mentioned in the Introduction, many grain surface models predict the formation of acetaldehyde via the coupling of HCO and $CH_3$. Our computations, though, predict that this process is very inefficient and, consequently, probably not responsible of the synthesis of acetaldehyde in the ISM. It remains the possibility, suggested by Ruaud et al. (2015), that acetaldehyde is synthesized by H-additions of CCO on icy grain surfaces. Alternatively, Charnley (2004) proposed that acetaldehyde is formed in gas phase by the oxidation of the ethyl radical ($C_2H_5$ + O → $CH_3CHO$ + H), where $C_2H_5$ is a daughter species of ethane ($C_2H_6$) or an intermediate product of the $C_2$ hydrogenation on the grains (e.g. Vasyunin & Herbst 2013). Model predictions in cold objects (Vasyunin & Herbst 2013; Vastel et al. 2014) and molecular shocks (Codella et al. 2015) support this gas-phase route of acetaldehyde formation. In summary, this work coupled with previous ones (Vasyunin & Herbst 2013; Vastel et al. 2014; Codella et al. 2015) tend to greatly diminish the role of reactions, other than hydrogenation, on the grain surfaces to synthesize acetaldehyde, and favour, on the contrary, a gas-phase route formation.

The second output of our computations is the binding energies of HCO and $CH_3$, 2333 and 734 K, respectively. These values are in rough agreement with the values used in the literature (Garrod et al. 2008 and subsequent works): 2200 and 1017 K, for HCO and $CH_3$, respectively.

The results presented in this Letter are of great relevance to our understanding of the formation of COMs, since other radical–radical pairs can show the same chemical behaviour. This is the case, for instance, for the reaction between two $CH_2OH$ radicals. Extrapolating our results to this case, the products in gas phase will be either $HOCH_2CH_2OH$ (ethylene glycol) as a consequence of the radical coupling, or $CH_3OH$ + $H_2CO$ due to H transfer, depending on the relative orientation of the radicals. However, on the surfaces of amorphous water ices, the second reactive channel will likely be the dominant one, contrarily to what suggested, for instance, by Garrod et al. (2008) and subsequent works (see Introduction; Chuang et al. 2016). A similar situation might also occur for the reaction between $CH_3O$ and $CH_3$, for which exclusive formation of $H_2CO$ + $CH_4$ is expected rather than $CH_3OCH_3$ (dimethyl ether), against, again, the assumption of the above mentioned grain-surface models. Of course, similar simulations as those carried out in this work are mandatory to have a firm answer whether the orientation of radicals on amorphous water ices play a dominant role or not in synthesizing COMs via the radical coupling mechanism.

A more general conclusion is that it is extremely important to have an atomic description of the reactions occurring on grain surfaces, as the interpretation of laboratory experiments and their extrapolation to astrochemical applications are extremely complex. Explicit accurate quantum chemical computations are mandatory to understand whether a given species can really be formed or not by the coupling of radicals on the grain surfaces, as chemical intuition could dramatically fail yielding erroneous products.

## ACKNOWLEDGEMENTS

AR is indebted to Programa Banco de Santander for a UAB distinguished postdoctoral research contract. Financial support from MINECO (projectsCTQ2015-62635-ERC and CTQ2014-60119-P) and DIUE (project 2014SGR482)are gratefully acknowledged. NB acknowledges the financial support from the UniversitÉ Joseph Fourier de Grenoble and the Observatoire de Grenoble, CC acknowledges the financial support from the French Space Agency CNES.